\documentclass[aps,prd,twocolumn,superscriptaddress]{revtex4-1}

\usepackage{tikz}
\usepackage{subcaption}
\usepackage{hyperref}   
\usepackage{graphicx}
\usepackage{epstopdf, epsfig}
\usepackage[french,english]{babel}
\usepackage{float}
\usepackage{amsmath}

\renewcommand{\vec}[1]{\boldsymbol{#1}}

\newcommand{\p}{\partial}

\newcommand{\be}{\begin{equation}}
\newcommand{\ee}{\end{equation}}
\newcommand{\bsub}{\begin{subequations}}
\newcommand{\esub}{\end{subequations}}
\newcommand{\bea}{\begin{eqnarray}}
\newcommand{\eea}{\end{eqnarray}}

\begin{document}

\title{Estimate of the superradiance spectrum in dispersive media}

\author{Theo Torres}
 \email{t.torres-vicente@sheffield.ac.uk}
\affiliation{Consortium for Fundamental Physics,
School of Mathematics and Statistics,
University of Sheffield, Hicks Building, Hounsfield Road, Sheffield S3 7RH, United Kingdom.}
\affiliation{
School of Mathematical Sciences, University of Nottingham, University Park, Nottingham, NG7
2RD, UK}

\begin{abstract}
In 2016, the Nottingham group detected the rotational superradiance effect. While this experiment demonstrated the robustness of the superradiance process, it still lacks a complete theoretical description due to the many effects at stage in the experiment. 
In this paper, we shine new light on this experiment by
deriving an estimate of the reflection coefficient in the dispersive regime by means of a WKB analysis.
This estimate is used to evaluate the reflection coefficient spectrum of counter rotating modes in the Nottingham experiment. Our finding suggests that the vortex flow in the superradiance experiment was not purely absorbing, contrary to the event
horizon of a rotating black hole. While this result increases the gap between this experimental vortex flow and a rotating black hole, it is argued that it is in fact this gap that is the source of novel ideas.
\end{abstract}
\maketitle

\section{Introduction}

In this contribution to \emph{the next generation of analogue gravity experiments} conference, we would like to focus on a previous analogue gravity experiment, the lessons learned from it and the ideas it gives us for the future. The experiment we will discuss here is the 2016 superradiance experiment conducted at the University of Nottingham presented in~\cite{Superradiance}.

Superradiance is a radiation enhancement process during which a wave scattering of a rotating object will extract some of the object's energy. 
This process was originally derived by Zel'Dovich by considering electromagnetic radiations incident on a rotating cylinder\cite{zeldovich1,zeldovich2}.
However, it was soon realised that this effect was not only restricted to the specific Zel'Dovich setup but can be applied to a broad class of systems~\cite{Brito:2015oca}.

This is because the superradiance effect does not rely on the underlying dynamical equations governing the system but rather on two conditions that can be satisfied in a large variety of physical situations.
The two fundamental conditions for superradiance to happen are~\cite{BRI15}:
\begin{itemize}
\item The system needs to allow for the presence of negative energy modes. When considering rotational superradiance, the scatterer must be a rotating object\footnote{We note that other amplification processes exist when considering waves scattering of a discontinuous media, known as over-reflection~\cite{Acheson76}, but we will focus on rotational superradiance here.}.
\item The system needs to provide an absorption mechanism.
\end{itemize}
One interesting system to satisfy these conditions are rotating black holes (BHs). 
Indeed, the ergosphere provides a mechanism to allow for the presence of negative energy and the event horizon acts as a perfect absorber. It was indeed shown, soon after Zel'Dovich's seminal paper, that rotating BHs can be the stage of the superradiance effect~\cite{misner,staro1,staro2}.
Since then, a long list of systems spanning the various fields of physics have been found to exhibit this effect, and some others are still being looked for. 
Amongst them are
water waves on a cylinder~\cite{Cardoso_detecting},
orbital angular momentum beam incident on a disk~\cite{Cisco18,gooding2019reinventing,Prain:2019jqk,faccio2019superradiant},
or binary BHs~\cite{Wong:2019kru}.

Following a long history of theoretical studies, the rotational superradiance effect was recently observed in the Nottingham experiment~\cite{Superradiance}. 
This experiment, motivated by the stunning fact that, \textit{under specific conditions}, waves propagating on a flowing fluid experience the presence of an effective curved space-time~\cite{Unruh:1980cg,SCH02}, was based on multiple theoretical studies of superradiance in analogue systems~\cite{basak,basak2,Slatyer:2005ty,Richartz:2009mi,Richartz:2012bd,Dolan_scattering,Richartz:2014lda}.
However, even with this whole array of studies, a complete theoretical description of the superradiance experiment is still missing. 
This is due to the various effects present in this experiment, the main ones being dispersion and vorticity, which even make the determination of effective governing equations a difficult task.
The effect of vorticity on shallow water flows has been studied in~\cite{flowmaster,Stepanyants18} while dispersive effects in irrotational flows have been investigated in~\cite{TheoLR}.
In this paper, built upon the work in~\cite{TheoLR}, we use a Wentzel-Kramers-Brillouin (WKB) approximation to predict the reflection coefficient spectrum to be compared with the experimental results of~\cite{Superradiance}. 
The paper is organised as follows: 
in section \ref{ray_sec}, we briefly review the work of \cite{TheoLR} in order to establish the notation and make the present study self-consistent. 
In section \ref{Tunnel_sec}, we derive the connection formula that relates WKB modes on both side of a saddle point. 
In section \ref{Estimate_sec}, we use the result of section \ref{Tunnel_sec} to compute the reflection coefficient spectrum in the presence of dispersive effect and we compare it with the experimental result of \cite{Superradiance}.
Finally, in section \ref{Conclusion_sec}, we discuss our findings and relate them to the next generation of analogue gravity experiments.

\section{Waves and rays}\label{ray_sec}

We consider here surface water waves propagating on top of a two-dimensional, axisymmetric, incompressible and irrotational vortex flow. The fluid velocity is given, in polar coordinates $(r,\theta)$, by the standard \textit{draining bathtub} (DBT) model~\cite{LandauV6}:
\begin{equation}\label{DBTflow}
\vec{v}_0 = \frac{C}{r}\vec{e}_\theta - \frac{D}{r}\vec{e}_r,
\end{equation}
where $C$ and $D$ are respectively the circulation and drain parameters. The surface of the water is assumed to be flat.
Such waves are irrotational perturbation to the background flow, they are therefore described by one scalar quantity, the velocity potential $\phi$, which obey the following equation of motion:
\begin{equation}\label{EoM}
\mathcal{D}_t^2\phi +F(-i\vec{\nabla})\phi = 0,
\end{equation}
where $\mathcal{D}_t = \partial_t + \vec{v}_0.\vec{\nabla}$ is the material derivative and $F$ contains the dispersion relation. 
In the following we will take the standard gravity waves dispersion relation, $F(k) = gk\tanh(h_0k)$, where $g$ is the gravitational acceleration and $h_0$ is the background fluid depth\footnote{Note that we neglect the effect of surface tension as its effects is negligible in the frequency range considered here}~\cite{Milewski96}.
Due to the spatial dependence of the background flow and the higher order derivatives involved in $F$,
 Eq.~\ref{EoM} cannot be solved exactly.
However, it is well known that solutions can be approximated by means of a gradient expansion method also known as WKB approximation~\cite{berry1972semiclassical}, we write:
\begin{equation}
\phi = A_{\mathrm{WKB}}e^{iS}
\end{equation}

At first order, the wave can be seen as a coherent collection of rays (see~\cite{synge1963hamiltonian} for an elegant presentation of the method for water waves). 
The rays are given by a Hamiltonian, which in our case is given by~\cite{TheoLR}:
\begin{equation}\label{Ham_def}
\mathcal{H} = \frac{1}{2}\left[(\omega - \vec{v}_0.\vec{k})^2 - F(\vec{k})\right],
\end{equation}
where $\vec{k} = \vec{\nabla}S = (k_r,m/r)$ and $\omega = -\partial_t S$, with $k_r$ being the radial wave-vector and $m$ being the azimuthal number.
The amplitude of the wave is calculated by using the fact that the wave action is transported with the group velocity, $v_g$, along tubes delimited by such rays. More precisely, in the absence of dissipation and in stationary systems, the amplitude satisfies the transport equation:
\begin{equation}
\vec{\nabla} \cdot ( \Omega_0 A_{\mathrm{WKB}}^{2} \vec{v}_g) = 0.
\end{equation}
Therefore, we have that:
\begin{equation}\label{transport_eq}
A_{\mathrm{WKB}} \propto \frac{1}{\sqrt{\Omega_0 v_g}},
\end{equation}
where $\Omega_0 = \omega - \vec{v}_0.\vec{k}$ is the co-moving frequency, computed after having solved Hamilton's equations with the Hamiltonian given in Eq.~\ref{Ham_def}.
 
\begin{figure}
\centering
\includegraphics[trim= 0.6cm 0 0 0,scale=0.7]{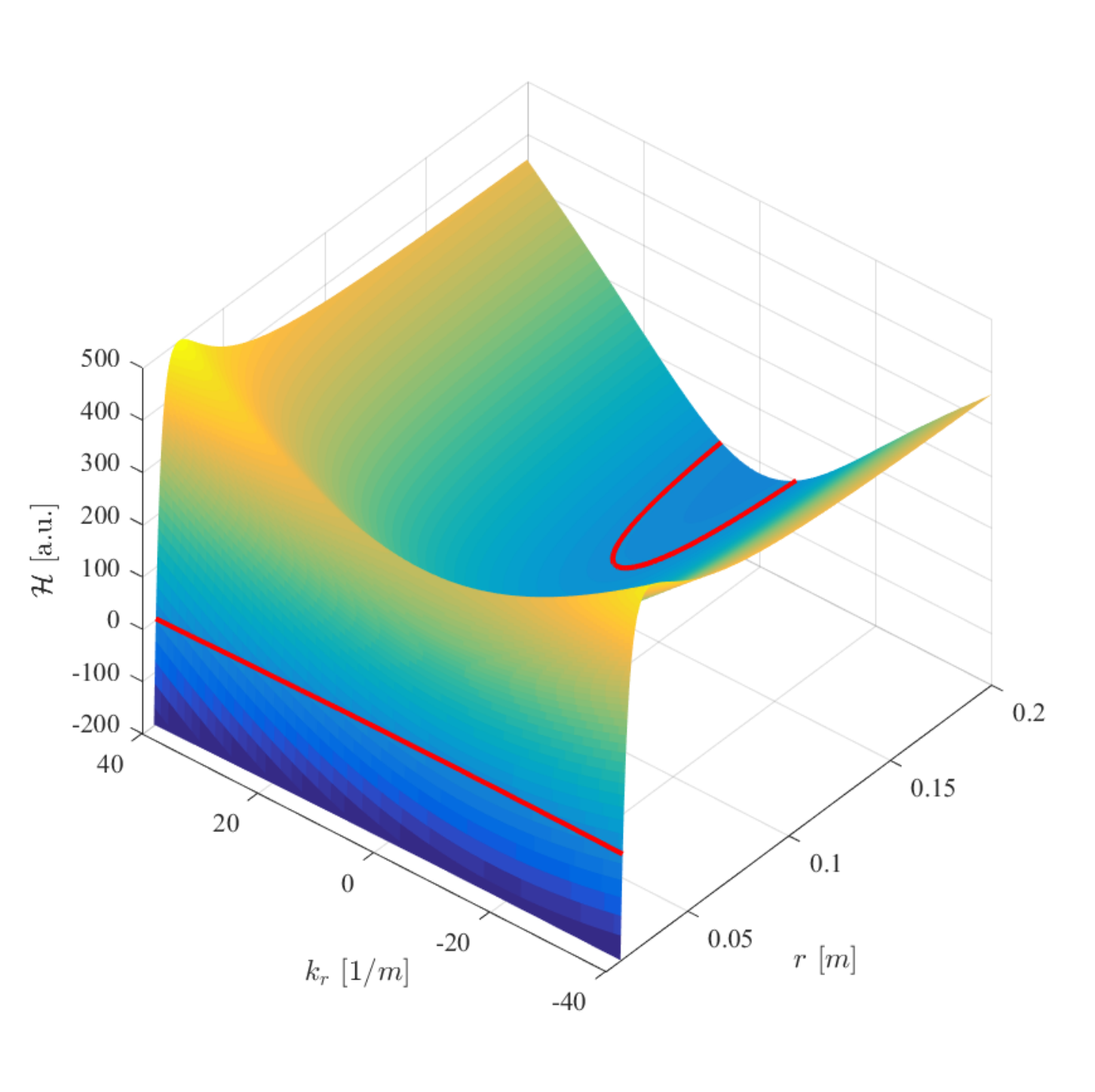}
\caption{Dispersion surface given by the Hamiltonian function $\mathcal{H}(r,k_r)$ in arbitrary units. The flow parameters are $C = 0.01~m^2/s$, $D = 0.001~m^2/s$ and $h_0 = 6~cm$. The surface is plotted for $m=1$ and a frequency $f=1~Hz$. 
The red lines correspond to the level line $\mathcal{H}(r,k_r) = 0$, i.e. waves satisfying the dispersion relation. We can see that the two lines are separated by a saddle point.}\label{dispersion_surface}
\end{figure}

At fixed $m$ and $\omega$ the Hamiltonian function defines a 2-dimensional surface parametrized by $r$ and $k_r$. 
The structure of the Hamiltonian function imposes that this surface admits a saddle point (see Fig.~\ref{dispersion_surface}). 

This is the approach developed in~\cite{TheoLR}, where the authors have used this approximation to construct the wavefronts of a wave incident on a DBT vortex flow with flow parameters corresponding to the Nottingham superradiance experiment. 
A good agreement was found between theory and experiment, at least sufficiently far from the vortex core where the flow can be considered irrotational. 
Furthermore, it was also shown in this framework that DBT vortices exhibit the analogue of black hole light-rings, even in the presence of dispersive effects. 
More precisely, pairs of $m$ and $\omega$ that are such that the Hamiltonian vanishes at the saddle point constitutes the light-ring spectrum. 
This light-ring spectrum is linked to the relaxation phase of vortices and was observed in a recent experiment~\cite{Torres_QNM}. 
The observed spectrum agrees with the theorical model developed in~\cite{TheoLR}. 
These two experimental verifications of the framework adopted in~\cite{TheoLR}, based on the DBT flow and the gravity waves dispersion relation, justify the use of Eqs.~\ref{DBTflow} and~\ref{EoM}. 

If one does not impose that the Hamiltonian vanishes at the saddle point, then the condition $\mathcal{H} = 0$ defines two disconnected region in phase space (see Fig.~\ref{dispersion_surface}), located on both sides of the saddle point. 
This structure allows for tunnelling in between the two regions.
It is possible to relate waves on each side of the saddle point (in the WKB limit) via a connection matrix $\mathcal{S}$ with coefficients determined by the saddle point as we show in the next section.
\section{Tunneling through a saddle point}\label{Tunnel_sec}

Here, we fix the value of $m$ and $\omega$ and assume that the Hamiltonian of the system $\mathcal{H}(r,k_r)$ has a saddle point at $(r^*,k_r^*)$. In the vicinity of the saddle point, the Hamiltonian can be expanded as:
\begin{equation}
H(r,k_r) = \mathcal{H}^* - \frac{1}{2} Y^T [\mathcal{H}] Y,
\end{equation}
where $\mathcal{H}^*$ is the value of the Hamiltonian evaluated at the saddle point, $Y$ is the vector $\binom{r - r^*}{ k_r - k_r^*}$ and $[\mathcal{H}]$ is the Hessian matrix of the Hamiltonian. 
 
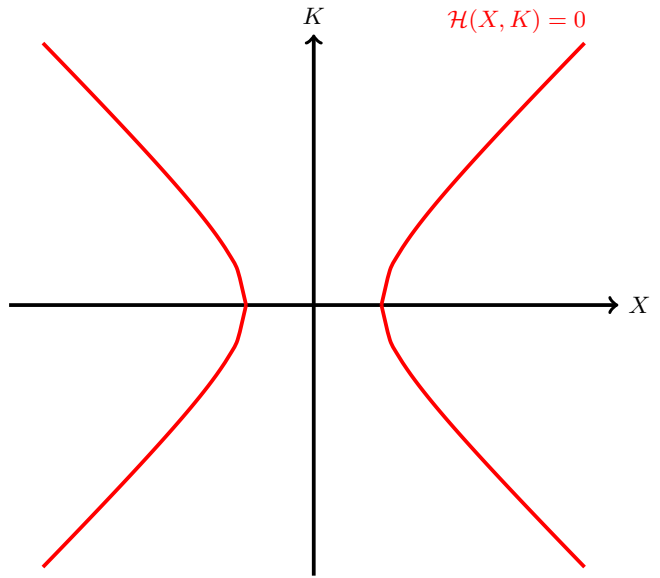
\begin{figure}[!h]
\centering
\begin{tikzpicture}[scale=0.9]
      \draw[line width=0.5mm,->] (-4.5,0) -- (4.5,0) node[right] {$X$};
      \draw[line width=0.5mm,->] (0,-4) -- (0,4) node[above] {$K$};
      \draw[line width=0.5mm,scale=1,domain=-4:-1,smooth,variable=\x,red] plot ({\x},{sqrt{abs{(\x*\x- 1}}});
      \draw[line width=0.5mm,scale=1,domain=-4:-1,smooth,variable=\x,red] plot ({\x},{-sqrt{abs{(\x*\x- 1}}});
      \draw[line width=0.5mm,scale=1,domain=1:4,smooth,variable=\x,red] plot ({\x},{sqrt{abs{(\x*\x- 1}}});
      \draw[line width=0.5mm,scale=1,domain=1:4,smooth,variable=\x,red] plot ({\x},{-sqrt{abs{(\x*\x- 1}}});
      
      \draw[red] (3,4.2) node {$\mathcal{H}(X,K) = 0$};

\end{tikzpicture}
\caption{Sketch of the phase-space diagram. The red curves represent the values of $X$ and $K$ satisfying the condition $\mathcal{H}(X,K)=0$ and correspond to the rays defined by the Hamiltonian $\mathcal{H}$. The saddle point is located at $(X,K)=(0,0)$}\label{phase_space_diag1}
    \end{figure}

Since $[\mathcal{H}]$ is a real and symmetric matrix, it can be diagonalised. 
We call $(\mu_1,\mu_2)$ the eigen-values and $(x,k)$ the components in the eigen-basis\footnote{In addition, the eigen-basis is orthogonal, which in two dimension is enough to ensure a canonical transformation $(r,k_r) \rightarrow (x,k)$.}.
The effective Hamiltonian becomes:
\begin{equation}
H = \mathcal{H}^* - \frac{\mu_1}{2} x^2 - \frac{\mu_2}{2} k^2.
\end{equation}
Finally performing the rescaling $X = (\mu_1/\mu_2)^{1/4}x$, which implies that $K = (\mu_2/\mu_1)^{1/4}k$, and dividing by $\sqrt{|\text{det}[\mathcal{H}|}$, we can bring the effective Hamiltonian to the form:
\begin{equation}\label{Ham}
 H(X,K) = \eta^2  - \frac{1}{2}(X^2 - K^2),
\end{equation}
with 
\begin{equation}
\eta^2 = \frac{\mathcal{H}^*}{\sqrt{|\text{det}[\mathcal{H}]|}}.
\end{equation}
The relative minus sign comes from the saddle structure, i.e. $\mu_1\mu_2<0$.
In phase-space this will appear as shown in Fig. \ref{phase_space_diag1}.

This effective Hamiltonian can be `lifted' at the level of a local wave equation \footnote{Note here that one needs to be careful when lifting a Hamiltonian to the level of a wave equation, as $x$ and $\p_x$ do not commute. This is rigorously done by means of the Weyl symbols~\cite{tracy2014ray}. However, the ordering is of no importance when working at the eikonal level.}:
\begin{equation}\label{local_we}
\left[ \eta^2  - \frac{1}{2}\left(X^2 + \frac{d^2}{dx^2}\right) \right] \phi = 0.
\end{equation}
Note that this equation describes tunneling through an
inverted harmonic oscillator potential.

In order to match WKB modes on both side of the saddle point, we need to find an exact solution of Eq.~\ref{local_we} valid globally. We then need to match the asymptotic expansions of this global solution to approximate WKB solutions sufficiently far away from the saddle point.

\subsection{Exact solutions}

We present here the global solution of Eq.~\ref{local_we} and the properties needed in order to perform the matching.

Eq.~\ref{local_we} can be solved exactly by means of parabolic cylinder function, $U$~\cite{Handbook}. Indeed, it admits as a general solution:
\begin{equation}\label{global_sol}
\phi(x) = A U(i \eta^2, \sqrt{2} X e^{-i\pi/4})  + B U( -i \eta^2, \sqrt{2} X e^{i\pi/4}).
\end{equation}
The function $U$ has the following asymptotic behaviour for $z\rightarrow \infty$:
\begin{equation}\label{asym1}
U(a,z) \approx e^{-z^2/4} z^{-a-1/2}   \quad \text{ for } \quad|arg(z)|<3\pi/4
\end{equation} 
\begin{eqnarray}\label{asym2}
U(a,z) \approx & & e^{-z^2/4} z^{-a-1/2} \nonumber\\
 &\pm & i\frac{\sqrt{2\pi}}{\Gamma(1/2 + a)} e^{\mp i \pi a} e^{z^2/4} z^{a-1/2}  \\ 
&\text{ for }& \quad \pi/4 < |arg(z)| <5\pi/4 \nonumber
\end{eqnarray}

We define the following functions to simplify the notation:
\begin{eqnarray}
U_1 &=& U\left(-i\eta^2, \sqrt{2}|X|e^{i\pi/4} \right), \\
U_2 &=& U\left(i\eta^2, \sqrt{2}|X|e^{-i\pi/4} \right), \\
U_3 &=& U\left(i\eta^2, -\sqrt{2}|X|e^{-i\pi/4} \right), \\
U_4 &=& U\left(-i\eta^2, -\sqrt{2}|X|e^{i\pi/4} \right),
\end{eqnarray}

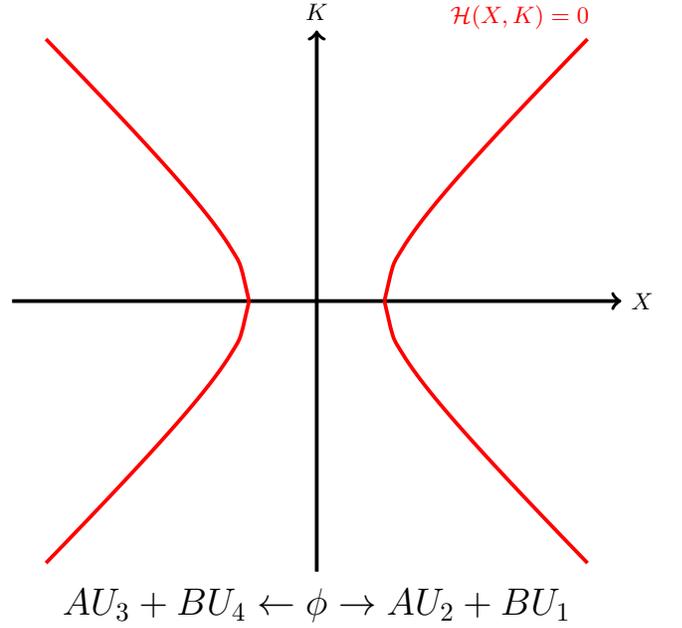
\begin{figure}
\centering
\begin{tikzpicture}[scale=0.9]
      \draw[line width=0.5mm,->] (-4.5,0) -- (4.5,0) node[right] {$X$};
      \draw[line width=0.5mm,->] (0,-4) -- (0,4) node[above] {$K$};
      \draw[line width=0.5mm,scale=1,domain=-4:-1,smooth,variable=\x,red] plot ({\x},{sqrt{abs{(\x*\x- 1}}});
      \draw[line width=0.5mm,scale=1,domain=-4:-1,smooth,variable=\x,red] plot ({\x},{-sqrt{abs{(\x*\x- 1}}});
      \draw[line width=0.5mm,scale=1,domain=1:4,smooth,variable=\x,red] plot ({\x},{sqrt{abs{(\x*\x- 1}}});
      \draw[line width=0.5mm,scale=1,domain=1:4,smooth,variable=\x,red] plot ({\x},{-sqrt{abs{(\x*\x- 1}}});
      
      \draw[red] (3,4.2) node {$\mathcal{H}(X,K) = 0$};
      \draw[black] (0,-4.5) node {\Large $ AU_3 + BU_4 \leftarrow \phi \rightarrow AU_2 + BU_1$};
\end{tikzpicture}
\caption{Phase-space diagram and asymptotic expansions of the solution to the local wave equation Eq.~\ref{local_we}.}\label{phase_space_diag2}

    \end{figure}

We can rewrite the solution given in Eq.~\ref{global_sol} as:
\begin{equation} 
\phi = 
     \begin{cases}
       AU_2 + BU_1 &\quad\text{if } X>0\\
       AU_3 + BU_4 &\quad\text{if } X<0 \\
     \end{cases}
\end{equation}
The asymptotic expansion are added to the phase-space diagram in Fig. \ref{phase_space_diag2}.
We now explicitly write down the asymptotic expansion of the functions $U_1$, $U_2$, $U_3$, and $U_4$.
The asymptotic expansion of $U_1$ and $U_2$ are obtained by setting $a = \pm i\eta^2$ and $z = \sqrt{2} |X| e^{\mp i \pi/4}$, and using the expansion given in Eq.~\ref{asym1} as $|arg(z)| = \pi/4$. We obtain:
\begin{equation}\label{global_exp1}
U_1 \approx \left( \sqrt{2} |X| \right)^{i\eta^2 - 1/2}\exp{\left[-iX^2/2\right]} \exp{\left[-\frac{\pi}{4}(\eta^2 + i/2)\right]},
\end{equation}
and
\begin{equation}\label{global_exp2}
U_2 \approx \left( \sqrt{2} |X| \right)^{-i\eta^2 - 1/2}\exp{\left[iX^2/2\right]} \exp{\left[-\frac{\pi}{4}(\eta^2 - i/2)\right]}.
\end{equation}

The expansion for $U_3$ and $U_4$ is obtained in the same manner with $a = \pm i\eta^2$ and $z = -\sqrt{2} |X| e^{\mp i\pi /4}$. Since $|arg(z)| = 3\pi / 4$, the asymptotic expansion given in Eq.~\ref{asym2} should be used. After some algebra, we obtain:
\begin{equation}\label{global_exp3}
U_3 \approx -i \tau^{-1} U_2 + \beta \tau^{-1} U_1,
\end{equation}
and
\begin{equation}\label{global_exp4}
U_4 \approx i \tau^{-1} U_1 + \beta^* \tau^{-1} U_2,
\end{equation}
where we have defined the following parameters:
\begin{equation}\label{tau_beta_def}
\tau = e^{-\pi \eta^2} \quad \text{and} \quad 
\beta = \frac{\sqrt{2\pi i \tau}}{\Gamma\left(i \eta^2 + 1/2 \right)}.
\end{equation}
Note that we have the following relation: $|\beta|^2 - \tau^2 =1$.
    
\subsection{WKB modes around the saddle point}
Here, we construct the WKB modes in a region where the quadratic approximation for the Hamiltonian is valid.
Since the WKB approximation breaks down at the saddle point, this construction must be done on both sides of the saddle point separately. 

We first start by computing the eikonal phase using the local Hamiltonian given in Eq.~\ref{Ham}. We have:
\begin{equation}
K = \p_X S = \pm \sqrt{X^2 - 2\eta^2}.
\end{equation}
After integration, we obtain the phase $S$:
\begin{eqnarray}
S(X) =& & S_0 \\
&\pm & \frac{1}{2} \left( X\sqrt{X^2 - 2\eta^2 } - 2\eta^2 \ln\left(\sqrt{X^2 - 2\eta^2} + X\right) \right), \nonumber
\end{eqnarray}
where $S_0$ is a constant of integration.
For $|X| \gg \eta$, this approximates to:
\begin{equation}\label{phase_WKB}
S(X) \approx \pm \left( \frac{X^2}{2} - \eta^2 \ln X \right) + S_0.
\end{equation}

We, then, compute the amplitude of the WKB mode. To do so, we first express the group velocity as:
\begin{equation}
v_g = \frac{dX}{d\sigma} = -\p_K H = K = \pm \sqrt{X^2 - 2\eta^2}.
\end{equation}
Inserting this result into the transport equation (Eq.~\ref{transport_eq}), we get that:
\begin{equation}
A_{\mathrm{WKB}} \propto \frac{1}{\left( X^2 - 2\eta^2\right)^{1/4}}.
\end{equation}
For $|X| \gg \eta$, this reduces to:
\begin{equation}\label{amp_WKB}
A_{\mathrm{WKB}} \approx |X|^{-1/2}.
\end{equation}
Combining Eqs.~\ref{amp_WKB} and~\ref{phase_WKB}, we obtain the WKB modes:
\begin{equation} \label{WKB_exp}
\phi_{\mathrm R,L}^{\rightleftharpoons} \propto  \big(|X| \big)^{\mp i \eta^2 - 1/2} \exp{\left[\pm i X^2/2 \right]}.
\end{equation}
The subscripts R and L denote that the solution is valid either to the right or to the left of the saddle point but not across. The superscript $\rightleftharpoons$ denotes the direction of propagation of the wave. To determine which direction corresponds to which sign, we look at the group velocity. 
If $v_g > 0$ the wave propagates towards the right, while if $v_g < 0 $ the wave propagates towards the left.
Therefore, we have explicitly the four WKB solutions. To the right of the saddle point, they are given by:
\begin{eqnarray}
\phi_{\mathrm R}^{\leftarrow} &\propto & \big( X \big)^{-i\eta^2 - 1/2} \exp{\left[ \frac{i X^2}{2} \right]} \label{WKB_R1}\\
\text{ and } \quad
\phi_{\mathrm R}^{\rightarrow} &\propto & \big( X \big)^{i\eta^2 - 1/2} \exp{\left[ \frac{-i X^2}{2} \right]}, \label{WKB_R2}
\end{eqnarray}
while to the left of the saddle point, they are given by:
\begin{eqnarray}
\phi_{\mathrm L}^{\rightarrow} &\propto & \big( |X| \big)^{-i\eta^2 - 1/2} \exp{\left[ \frac{i X^2}{2} \right]}
\quad \label{WKB_L1}\\
\text{ and } \quad
\phi_{\mathrm L}^{\leftarrow} &\propto & \big( |X| \big)^{i\eta^2 - 1/2} \exp{\left[ \frac{-i X^2}{2} \right]}.\label{WKB_L2}
\end{eqnarray}

\begin{figure}
\centering
\begin{tikzpicture}[scale=0.9]
      \draw[line width=0.5mm,->] (-4.5,0) -- (4.5,0) node[right] {$X$};
      \draw[line width=0.5mm,->] (0,-4) -- (0,4) node[above] {$K$};
      \draw[line width=0.5mm,scale=1,domain=-4:-1,smooth,variable=\x,red] plot ({\x},{sqrt{abs{(\x*\x- 1}}});
      \draw[line width=0.5mm,scale=1,domain=-4:-1,smooth,variable=\x,red] plot ({\x},{-sqrt{abs{(\x*\x- 1}}});
      \draw[line width=0.5mm,scale=1,domain=1:4,smooth,variable=\x,red] plot ({\x},{sqrt{abs{(\x*\x- 1}}});
      \draw[line width=0.5mm,scale=1,domain=1:4,smooth,variable=\x,red] plot ({\x},{-sqrt{abs{(\x*\x- 1}}});
      \draw[red] (3,4.2) node {$\mathcal{H}(X,K) = 0$};
      \draw[black] (0,-4.5) node {\Large $ AU_3 + BU_4 \leftarrow \phi \rightarrow AU_2 + BU_1$};
      \draw[line width=0.5mm] (-2,1.73-0.2) -- (-2,1.73) -- (-2+0.2, 1.73);
      \draw[black] (-3, 1.73) node {\Large $\phi_{\mathrm L}^{\leftarrow}$};
      
      \draw[line width=0.5mm] (-2-0.2,-1.73) -- (-2,-1.73) -- (-2, -1.73 -0.2);
      \draw[black] (-3, -1.73) node {\Large $\phi_{\mathrm L}^{\rightarrow}$};
      
      \draw[line width=0.5mm] (2 -0.2,-1.73) -- (2,-1.73) -- (2, -1.73 + 0.2);
      \draw[black] (3, -1.73) node {\Large $\phi_{\mathrm R}^{\rightarrow}$};
      
      \draw[line width=0.5mm] (2,1.73+0.2) -- (2,1.73) -- (2+0.2, 1.73);
      \draw[black] (3, 1.73) node {\Large $\phi_{\mathrm R}^{\leftarrow}$};
      \draw[black] (2.7,-5.5) node {\Large $\rightarrow a_{\mathrm R}^{\leftarrow}\phi_{\mathrm R}^{\leftarrow} +
a_{\mathrm R}^{\rightarrow}\phi_{\mathrm R}^{\rightarrow} $};
      \draw[black] (-2.7,-5.5) node {\Large $ a_{\mathrm L}^{\leftarrow}\phi_{\mathrm L}^{\leftarrow} +
a_{\mathrm L}^{\rightarrow}\phi_{\mathrm L}^{\rightarrow} \leftarrow $};

\end{tikzpicture}
\caption{Phase-space diagram and WKB modes on both sides of the saddle point. The arrow on the red curves indicate the direction of propagation of the WKB modes.}\label{phase_space_diag3}
    \end{figure}
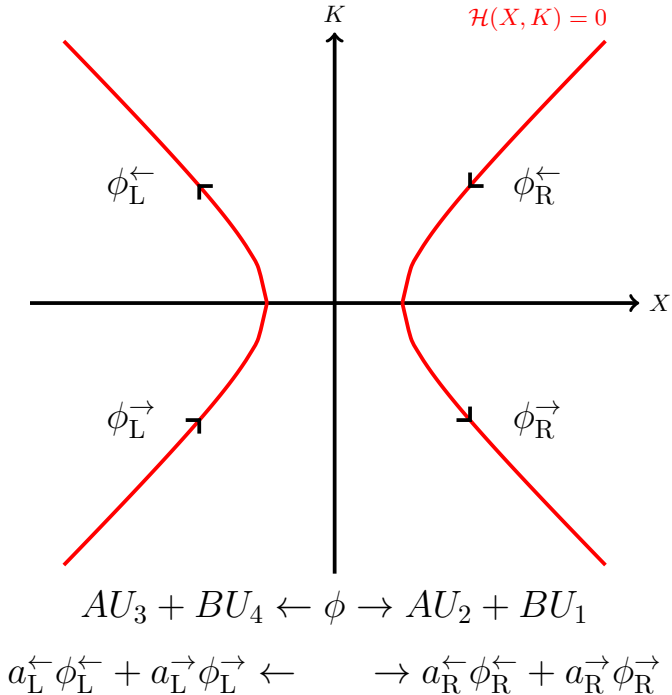

The WKB modes are added to the phase-space diagram in Fig. \ref{phase_space_diag3}.

    \subsection{Matching conditions}
    
    We have solved, both exactly and by means of a WKB expansion, the local wave equation (Eq.~\ref{local_we}). The two solutions should match sufficiently far away from the saddle point as can be seen from the asymptotic expansions of both solutions. We now work out the matching conditions and relate the WKB modes on both sides of the saddle point.
    
    First we need to identify the correspondence between the asymptotic expansion of the exact solutions $U_j$ and the WKB modes $\phi_{\mathrm R,L}^{\rightleftharpoons}$.
    
    For $X\gg \eta$, using the asymptotic expressions given in Eqs.~\ref{global_exp1}, \ref{global_exp2}, as well as Eqs.~\ref{WKB_R1}, and \ref{WKB_R2} ,we have that:
\begin{eqnarray}
U_1 &=& \alpha \gamma \phi_{\mathrm R}^{\rightarrow} 
\quad \label{match_R1}\\
 \text{and} \quad 
U_2 &=& \alpha^{-1} \gamma \phi_{\mathrm R}^{\leftarrow}.\label{match_R2}
\end{eqnarray}
where the parameters $\alpha$ and $\gamma$ are defined as:
\begin{equation} \label{alpha_def}
\alpha = \sqrt{2}^{i\eta^2} e^{-i\pi /8} 
\quad \text{and} \quad
\gamma = 2^{-1/4} e^{-\pi \eta^2 /4}.
\end{equation}

Now using Eqs.~\ref{global_exp3} and~\ref{global_exp4}, in addition to Eqs.~\ref{WKB_L1} and \ref{WKB_L2}, we obtain:
\begin{eqnarray}
U_3 &=& -i \tau^{-1} \alpha^{-1} \gamma \phi_{\mathrm L}^{\rightarrow} + \beta\tau^{-1} \alpha \gamma \phi_{\mathrm L}^{\leftarrow}
\quad \label{match_L1}\\
\text{and} \quad
U_4 &=& i\tau^{-1}\alpha\gamma\phi_{\mathrm L}^{\leftarrow} + \beta^*\tau^{-1} \alpha^{-1} \gamma \phi_{\mathrm L}^{\rightarrow}. \label{match_L2}
\end{eqnarray}

Eqs.~\ref{match_R1} and \ref{match_R2} give the matching conditions to the right of the saddle point. It allows us to relate the amplitude of the right WKB modes with the amplitude of the exact solution. This leads to:
\begin{equation}
A = \alpha \gamma^{-1} a_{\mathrm R}^{\leftarrow} 
\quad \text{and} \quad
B = \alpha^{-1} \gamma^{-1} a_{\mathrm R}^{\rightarrow}.
\end{equation}

The same procedure using Eqs.~\ref{match_L1} and~\ref{match_L2} allows us to connect $a_{\mathrm L}^{\rightleftharpoons}$ with $A$ and $B$:
\begin{eqnarray}
a_{\mathrm L}^{\rightarrow} &=& A \beta \tau^{-1} \alpha \gamma + B i \tau^{-1} \alpha \gamma
\\
a_{\mathrm L}^{\leftarrow} &=& -A i \tau^{-1} \alpha^{-1} \gamma + B \beta^* \tau^{-1} \alpha^{-1} \gamma.
\end{eqnarray}

Combining the previous relations, we finally obtain the connection formula between the right and left WKB mode amplitudes:
\begin{equation}\label{match_SP}
\binom{a_{\mathrm L}^{\leftarrow}}{a_{\mathrm L}^{\rightarrow}} = 
\mathcal{S}  
\binom{a_{\mathrm R}^{\leftarrow}}{a_{\mathrm R}^{\rightarrow}}
\end{equation}
where $\mathcal{S}$, which is fully determined by the value of the Hamiltonian on the saddle point and by the value of the determinant of the Hessian matrix, is given by:
\begin{equation}\label{connection_matrix}
\mathcal{S}=
  \begin{bmatrix}
      \beta \tau^{-1} \alpha^{2} & i\tau^{-1} \\
    -i \tau^{-1} & \beta^*\tau^{-1}\alpha^{-2} \\
  \end{bmatrix}.
\end{equation}
\section{Estimate of the reflection coefficient}\label{Estimate_sec}

We can now estimate the reflection coefficient spectrum by imposing boundary conditions at the vortex core in the connection formula \ref{match_SP}. 
Once the boundary conditions are imposed, we find, for different frequencies, the entries of the matrix $\mathcal{S}$ and we solve Eq.\ref{absorber} to extract the value of the reflection coefficient.

We first note that for the DBT flow parameters of the Nottingham experiment, this approach is only valid for $m<0$. 
Indeed, for $m>0$, we find that the saddle point of the Hamiltonian (which determines the scattering) is located inside the vortex core, in a region where the DBT model fails to represent the fluid flow.
However, for $m<0$, this saddle point is located outside the vortex and the flow can be well approximated by the DBT model.
In the following, we will therefore focus on the counter-rotating modes $m=-1$ and $m=-2$, and we choose the flow parameters to be $C = 0.016~m.s^{-1}$ and $D =0.001~m.s^{-1}$.

In analogy with a rotating black hole, we first assume that the vortex is purely absorbing and that waves are only falling in the hole. This is represented by the following connection
formula:
\begin{equation}\label{absorber}
\binom{T}{0} = 
\mathcal{S}  
\binom{1}{R},
\end{equation}
where $R$ and $T$ are respectively the reflection and transmission coefficient. 
 
\begin{figure}[!h]
\centering\includegraphics[trim=1cm 0 0 0, scale=0.85]{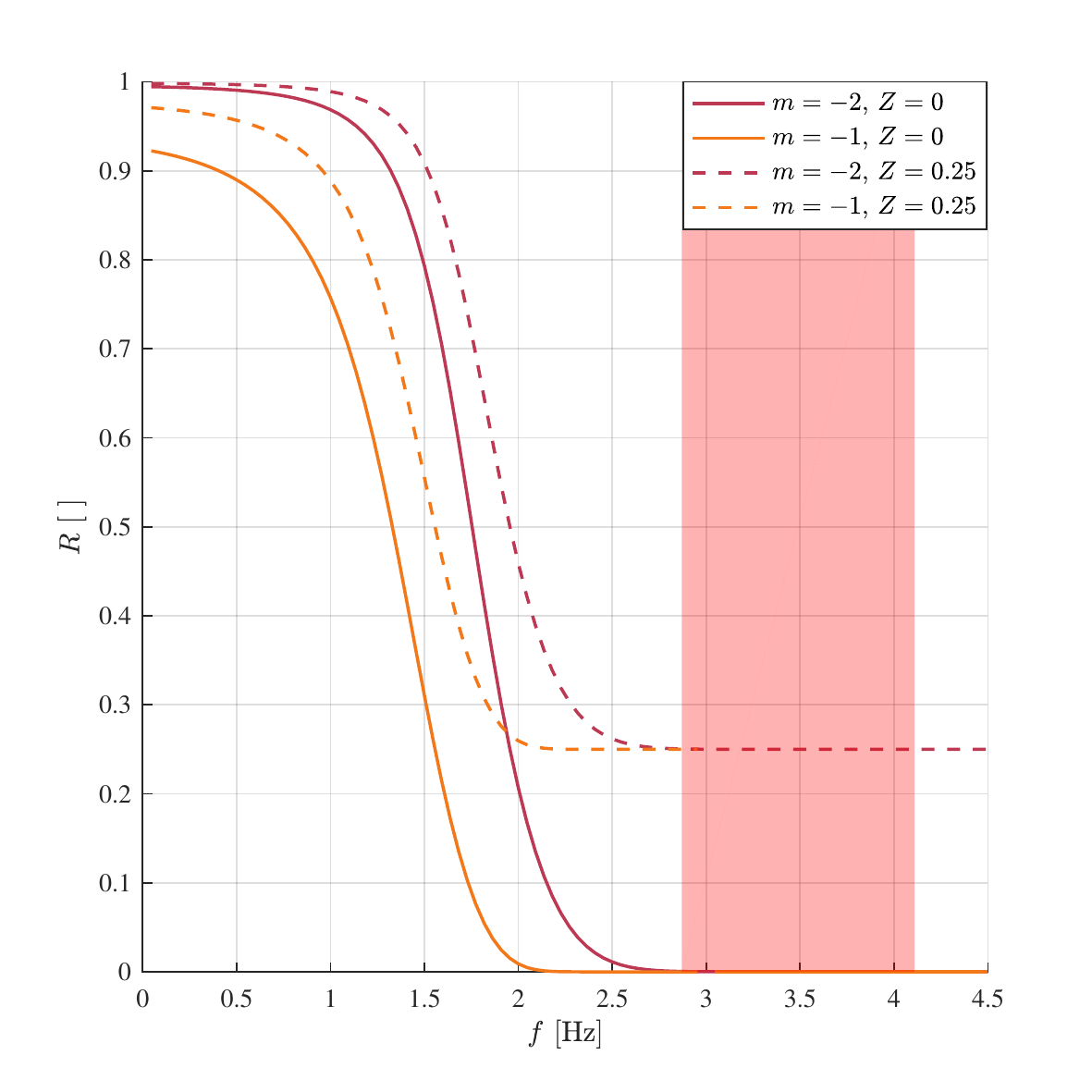}
\caption{WKB estimate of the reflection coefficient spectrum of the counter rotating modes, $m=-1$ in orange and $m=-2$ in purple, for the flow parameters corresponding to the Nottingham experiment~\cite{Superradiance}. 
The solid curves depict the reflection coefficient when the vortex is supposed to be purely absorbing, similar a rotating black hole, i.e. $Z=0$. 
The dashed curves represent the reflection coefficient spectra for a non-purely absorbing vortex, i.e. $Z \neq 0$. The frequency range corresponding to the regime explored in~\cite{Superradiance} is delimited by the red region. We can see that for these parameters, the reflection coefficient is essentially vanishing for a purely absorbing vortex while it asymptotes to a constant value (corresponding to the chosen value of $Z$ for a non-purely absorbing vortex.}
\label{fig_sim}
\end{figure}

The WKB spectrum obtained is presented by the solid curved in \ref{fig_sim}.
We can see that in the frequency range of the Nottingham experiment, represented by the faded red rectangle in Fig.\ref{fig_sim}, the reflection coefficient for $m = -1$ and $m=-2$ essentially vanishes. 
This behaviour differs from the experimental spectrum. 
Indeed, in Fig.2 of \cite{Superradiance}, we see that the reflection coefficient of the counter-rotating modes asymptotes a constant value. 
This value depends on the azimuthal number and is about $0.2$ for $m=-1$ and smaller than $0.1$ for $m=-2$.

This observation can be explained if one does not consider the vortex core to be a purely absorbing membrane, as a rotating black hole, but rather reflects parts of the waves.
We therefore modify the connection formula and assume that the vortex core has an impedance characterised by the coefficient $Z$:
\begin{equation}\label{reflector}
\binom{(1-Z)T}{ZT} = 
\hat{\mathcal{S}}
\binom{1}{R},
\end{equation}
where $\hat{\mathcal{S}}$ denotes the fact that we have taken into account the extra reflections between the reflected wave from the vortex core and the saddle point.
The spectra obtained for a non-purely absorbing vortex are represented by the dashed curved in Fig.\ref{fig_sim} for $Z=0.25$. 
The value of $Z$ is arbitrary and is chosen only to illustrate the result.
We can see that when we consider a reflecting vortex core, the reflection coefficient spectrum does not vanish at high frequency but rather asymptotes a constant value, determined by the value of $Z$.

This observation suggests that the vortex flow of the Nottingham was not a purely absorbing boundary in the manner of a rotating black hole. 
From the experimental spectrum of \cite{Superradiance}, we can also infer the value of the reflection at the vortex core for the various azimuthal numbers. 
In particular, we read that $Z_{m=-1} \approx 0.2$ and $Z_{m=-2} \approx 0.07$.

\section{Conclusion}\label{Conclusion_sec}

In this paper, we have studied the scattering of surface waves on top of an irrotational vortex flow and we have included the effect of dispersion in a WKB estimate of the reflection coefficient. This is done in order to match the experimental spectrum observed in the Nottingham experiment~\cite{Superradiance}. 
We have seen that this approach is not applicable to compute the spectrum of co-rotating waves. 
This is because the connection of WKB modes must be done at a radius at which vorticity effects need to be considered as well. 
However, this method should be valid to estimate the spectrum for the counter-rotating wave. 
If we are to believe this estimate, we must conclude that the vortex core of the Nottingham experiment does not behave as a purely absorbing boundary and that some reflection at the vortex core must be included.
While the explanation behind a reflection at the drain is still unclear, the presence of such a mechanism is to be expected. 
Indeed, as already mentioned, we know that the flow becomes more complicated than the DBT model close to the centre.
A more complicated background fluid flow will result in new phenomena involved in the scattering process. 
For example, it has been shown that the presence of vorticity on shallow water waves introduces extra structure to the Hamiltonian (or equivalently, to the effective potential)~\cite{flowmaster}. 
In particular, this introduces an extra maximum to the potential on which waves can scatter.
This new maximum is located between the usual maximum due to the irrotational flow and the drain.
Combining this extra bump with a purely absorbing drain can appear as a new not-perfectly absorbing effective boundary. 
Moreover, since the shape of the potential depends on the azimuthal number, the value of the effective reflection could depend on the value of $m$. 
In addition to vorticity effects, we also know that the height of the water changes closer to the drain. 
The effect of such a change is still unclear and could play a role in modifying the inner boundary.
It is important to note that even though extra structures are present, the effect of modified boundary conditions does not prevent the superradiance effect to occur~\cite{Richartz:2009mi}. Exploring these effects both theoretically and experimentally is an interesting open challenge.

The Nottingham experiment was motivated by the analogy between surface waves on a flowing fluid and fields in curved space-time. 
However, it is known that this analogy holds only under specific conditions, which were not all satisfied in this experiment.
In particular, it was clear that vorticity effects were significant closer to the vortex core and that dispersive effects could not be neglected, implying that a direct analogy between the vortex flow of the Nottingham experiment and a rotating black hole was impossible. 
In this paper, we have somehow increased this gap by showing that the vortex flow was not a perfect absorber, unlike a black hole.
In our view, this gap makes the observation of superradiance in the Nottingham experiment even more interesting as it strengthened the robustness of the effect but also opened new questions to answer.
For example, including dispersive effects at the eikonal level led to the development of the idea of light-rings for fluid flows~\cite{TheoLR} and motivated an experiment to observe them~\cite{Torres_QNM}. 
Inspired by the fluid-gravity analogy and the idea of spectroscopy, the observation of analogue light-rings stimulated the development of a new flow measurement technique applicable to fluids and superfluids alike~\cite{Torres_ABHS}.
Similarly, the Nottingham experiment sparked the interest to study the effect of vorticity in wave/vortex scattering~\cite{flowmaster,Stepanyants18} and revealed in particular that rotational vortex flows can exhibit the presence of quasi-bound states which remains to be seen.
Finally, the superradiance experiment motivated the study of backreaction in an analogue system~\cite{backreaction}.

It is clear from this example that the deviations from the analogue regime (vorticity, dispersion, etc...) have spurred the development of our concepts to new regimes which deepened our understanding of various processes such as superradiance or vortex relaxation.
These extra effects, inherently present in any experiment, compelled us to extend our ideas and are the sources of new knowledge.
We therefore believe that performing analogue gravity experiments that include these extra features, instead of suppressing them, will stimulate new problems to solve and will result in advances not only in the field of analogue gravity, but also in condensed matter and gravitational physics.
Here lies the strength of analogue gravity, in its capacity to stimulate new ideas, and to develop intuitions and feelings about the various ways nature expresses itself.



\acknowledgements
With many thanks to Antonin Coutant, Sam Patrick and Silke Weinfurtner for the many discussions that led to that study.


%
%
%
%
%
\bibliographystyle{utphys}
\bibliography{bibliography_PhD}

\end{document}